\documentclass[apj]{emulateapj}
\usepackage{mathptmx}

\newcommand  \acc     {\ifmmode {\rm km\,s}^{-2} \else km\,s$^{-2}$\fi}
\newcommand  \kms      {\ifmmode {\rm km\,s}^{-1} \else km\,s$^{-1}$\fi}

\newcommand  \cmii     {\hbox{cm$^{-2}$}}
\newcommand  \ergs     {\ifmmode {\rm erg\,s}^{-1} \else erg s$^{-1}$\fi}
\newcommand  \ergcms   {\ifmmode {\rm erg\,cm}^{-2}\,{\rm s}^{-1}
                        \else erg\,cm$^{-2}$\,s$^{-1}$\fi}
\newcommand  \ergcmsA  {\ifmmode{\rm erg\,cm}^{-2}\,{\rm s}^{-1}\,{\rm\AA}^{-1}
                        \else erg\,cm$^{-2}$\,s$^{-1}$\,\AA$^{-1}$\fi}
\newcommand  \ergcmsHz {\ifmmode{\rm erg\,cm}^{-2}\,{\rm s}^{-1}\,{\rm Hz}^{-1}
                        \else erg\,cm$^{-2}$\,s$^{-1}$\,Hz$^{-1}$\fi}
\newcommand  \phcms    {\ifmmode {\rm ph\,cm}^{-2}\,{\rm s}^{-1}
                        \else ph\,cm$^{-2}$\,s$^{-1}$\fi}
\newcommand  \phcmsA   {\ifmmode {\rm ph\,cm}^{-2}\,{\rm s}^{-1}\,{\rm\AA}^{-1}
                        \else ph\,cm$^{-2}$\,s$^{-1}$\,\AA$^{-1}$\fi}

\journalinfo{The Astronomical Journal, 
127: ??1--??9, 2004 May, astro-ph/0402318}
\slugcomment{Received 2003 December 12; accepted 2004 February 11}

\shorttitle{FUSE SPECTROSCOPY OF NGC\,4051}

\shortauthors{KASPI ET AL.}

\begin{document}

\title{{\it Far Ultraviolet Spectroscopic Explorer} Spectroscopy of Absorption and Emission Lines \\ from the Narrow-Line Seyfert 1 Galaxy NGC\,4051}
\author{
Shai~Kaspi,\altaffilmark{1,2} 
W.~N.~Brandt,\altaffilmark{3}
M.~J.~Collinge,\altaffilmark{4}
Martin Elvis,\altaffilmark{5}
and
Christopher S. Reynolds\altaffilmark{6}
}

\altaffiltext{1}{School of Physics and Astronomy, Raymond and Beverly
Sackler Faculty of Exact Sciences, Tel-Aviv University, Tel-Aviv 69978,
Israel; shai@wise.tau.ac.il}
\altaffiltext{2}{Physics Department, Technion, Haifa 32000, Israel.}
\altaffiltext{3}{Department of Astronomy \& Astrophysics, 525 Davey
Laboratory, The Pennsylvania State University, University Park, PA 16802, USA.}
\altaffiltext{4}{Princeton University Observatory, Princeton, NJ 08544, USA.}
\altaffiltext{5}{Harvard-Smithsonian Center for Astrophysics, 60 Garden Street, Cambridge, MA 02138, USA.}
\altaffiltext{6}{Department of Astronomy, University of Maryland,
College Park, MD 20742, USA.}

\begin{abstract}

We present three {\it Far Ultraviolet Spectroscopic Explorer (FUSE)\/}
observations of the Narrow-Line Seyfert~1 galaxy NGC\,4051. The
most prominent features in the far-ultraviolet (FUV) spectrum are
the \ion{O}{6} emission and absorption lines and the \ion{H}{1}
Lyman series absorption lines which are detected up to the Lyman
edge. We also identify weak emission from \ion{N}{3}, \ion{C}{3},
and \ion{He}{2}. The \ion{C}{3} line shows absorption while none
is detected in the \ion{N}{3} and \ion{He}{2} lines.  In \ion{H}{1}
and \ion{C}{3} we detect two main absorption systems at outflow
velocities of $-50\pm30$ and $-240\pm40$ \kms , as well as a possible
third one at $\sim -450$ \kms . These systems are consistent in
velocity with the 10 absorption systems found previously in \ion{C}{4},
\ion{N}{5}, and \ion{Si}{4}, though the individual systems are blended
together in the FUV spectrum. We estimate column densities of the
two main absorption systems and find that the \ion{H}{1} column density
is lower for systems with larger outflow velocity. We detect no flux
or spectral variations of NGC\,4051 at FUV wavelengths during three
epochs spanning one year. This is consistent with the optical light
curve which shows no variations between the three epochs. It is also
consistent with the X-ray light curve which shows consistent flux
levels at the three epochs of the {\it FUSE\/} observations, although
the X-ray light curve shows strong variations on much shorter timescales.

\end{abstract}

\keywords{
galaxies: active --- 
galaxies: individual (NGC\,4051) --- 
galaxies: nuclei --- 
galaxies: Seyfert --- 
techniques: spectroscopic ---
ultraviolet: galaxies}

\section{Introduction}

The Narrow-Line Seyfert~1 (NLS1) galaxy NGC\,4051 is one of the
best-studied active galactic nuclei (AGN) across the electromagnetic
spectrum due to its brightness and proximity ($V\approx 13.5$;
$z=0.002295\pm 0.000043$ from optical emission lines; de Vaucouleurs
et~al. 1991). It has been part of many studies since it was first
mentioned by Hubble (1926) and classified by Seyfert (1943). Collinge
et~al. (2001) recently presented the first high-resolution X-ray and
ultraviolet (UV) spectra of NGC\,4051. The {\it Chandra\/} High-Energy
Transmission Grating Spectrometer (HETGS) spectrum reveals absorption
and emission lines from hydrogen-like and helium-like ions of O, Ne,
Mg, and Si. Two distinct blueshifted absorption systems were detected:
a high-velocity system at $-2340\pm130$ \kms\ and a low-velocity system
at $-600\pm130$ \kms. In a {\it Chandra\/} Low-Energy Transmission
Grating Spectrometer (LETGS) observation taken $\sim 2$~yr later, van
der Meer et~al. (2003) claim to detect a new X-ray absorption system at
$\sim -4500$ \kms\ while the $-2340$ \kms\ system is barely visible.
An even higher velocity outflowing absorption system at $\sim -6500$
\kms\ or $\sim -16500$ \kms\ (depending on the identification of an
absorption line at 7.1 keV in an {\it XMM-Newton} EPIC observation)
was suggested recently by Pounds et al. (2003). The UV spectrum
taken with the Space Telescope Imaging Spectrograph (STIS) shows
strong absorption mainly in the lines of \ion{C}{4}, \ion{N}{5},
and \ion{Si}{4}. Ten intrinsic UV absorption systems are seen with
velocities between $-650$ and 30 \kms\ and FWHMs ranging from $\sim 30$
to $\sim 160$ \kms .  While the low-velocity X-ray absorption system
is consistent in velocity with many of the UV absorption systems,
the high-velocity X-ray absorption seems to have no UV counterpart.

NGC\,4051 shows rapid and large-amplitude X-ray flux variability by
up to an order of magnitude on timescales of hours (e.g., Marshall
et al. 1983; Uttley et~al. 2000; Collinge et~al. 2001; McHardy et
al. 2003; and references therein). Uttley et~al. (2000) found a strong
correlation between the variability in the extreme UV (124--188~eV)
and X-ray (2--10~keV) bands, and they suggested this to indicate
that both bands are sampling the same power-law continuum. 
Shemmer et~al. (2003) find the variable optical continuum flux to
correlate with the variable X-ray flux. The optical continuum leads
the X-ray by $2.4\pm1.0$ days. They also found that the variation
amplitude in the optical is much smaller than in the X-ray, as seen
in other NLS1s.

In this paper we bridge between the X-ray and near UV high-resolution
spectra by presenting the first {\it Far Ultraviolet Spectroscopic
Explorer (FUSE)\/} spectra of NGC\,4051.  The {\it FUSE\/} spectra
cover the 900--1180\,\AA\ band with a resolution of $\sim 0.06$\,\AA\
(corresponding to $\sim 20$~km~s$^{-1}$) . Our main proposed goals were
(1) to study the FUV kinematic counterparts to the known X-ray and UV
absorption lines, 
(2) to study any FUV absorption-line variability, and 
(3) to relate any FUV variability to that seen at X-ray and other wavelengths.  
The observations and data reduction are detailed in \S~2. In \S\S~3 and
4 we present the temporal and spectral analyses, and in \S~5 we discuss
our results.  Finally, \S6 presents a summary of our main results.

\section{Observations and Data Reduction}

NGC\,4051 was observed with {\it FUSE\/} (Moos et~al. 2000; Sahnow et
al. 2000) during three epochs, spanning one year, which are listed
in Table~\ref{obsinfo}.  The observations were carried out using
the $30\arcsec \times 30\arcsec$ low-resolution (LWRS) aperture in
the standard observing mode. Each observation consists of several
consecutive orbits, and in each orbit data from eight detectors (each
covering a different wavelength range) were accumulated.  We reduced
the raw data using the most recent {\it FUSE\/} software (CalFUSE
v2.4.0 and the {\it FUSE\/} IDL tools version of 2002 July).  For each
observation we checked the data from different detectors and orbits
for consistency and found no significant deviations in flux. Hence,
we used the FUSE\_REGISTER tool to combine the data sets from the
different detectors and orbits into one spectrum per observation.

In the following analysis we use the line templates from Feldman
et~al. (2001) to identify the airglow geocoronal emission lines and
Sembach (1999) to identify interstellar Galactic absorption lines.

\begin{deluxetable}{cccc}
\tablecolumns{4}
\tablewidth{0pt}
\tablecaption{{\em FUSE} Observation Log for NGC\,4051
\label{obsinfo}}
\tablehead{
\colhead{Obs. ID} &
\colhead{Date and UT} &
\colhead{$N_{\rm orbits}$\tablenotemark{a}} &
\colhead{Time (ks)\tablenotemark{b}}
 }
\startdata
B0620201000 & 2002 Mar 29 01:58 & 21 & 28.7  \\
C0190101000 & 2003 Jan 18 11:43 & 16 & 34.7  \\
C0190102000 & 2003 Mar 19 07:46 & 17 & \phn28.8  
\enddata
\tablenotetext{a}{Number of consecutive orbits.}
\tablenotetext{b}{Total effective exposure time.}
\end{deluxetable}

\section{Temporal Analysis}

\begin{figure*}
\centerline{\includegraphics[width=18.0cm]{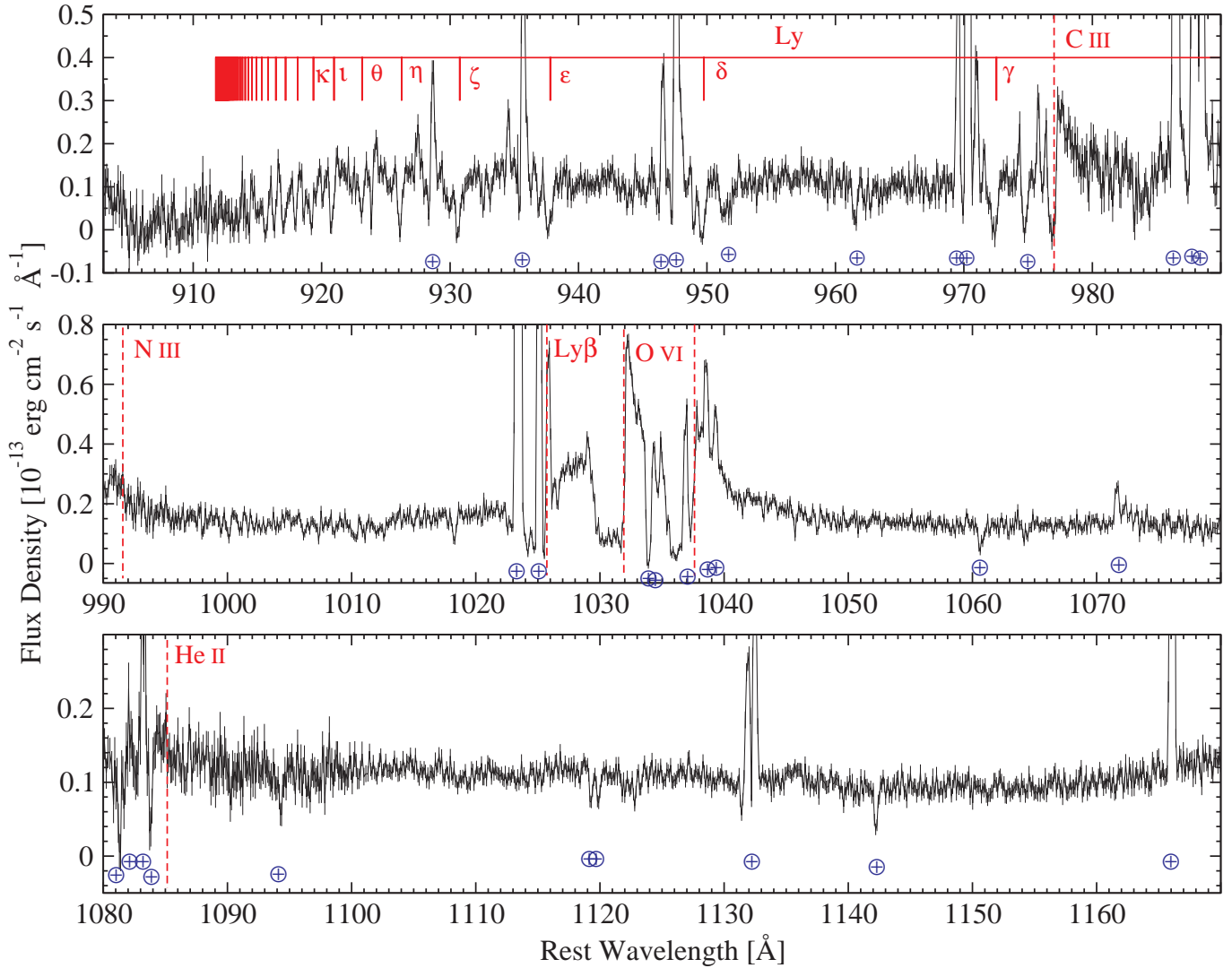}}
\caption{The combined 92.2 ks {\it FUSE\/} spectrum of NGC\,4051 binned
to $\sim 0.1$ \AA~pix$^{-1}$.  Intrinsic emission lines are marked
at their theoretically expected positions in red above the spectrum.
Airglow geocoronal emission lines and interstellar Galactic absorption
lines are marked with~$\oplus$.  The intrinsic \ion{H}{1} absorption
lines are marked at their theoretically expected positions up to the
ionization energy in the top panel.
\label{fusespec} }
\vglue -0.1cm
\end{figure*}

Visual comparison of the spectra from the three epochs suggests
striking similarity between them with no apparent time variation of
the intrinsic spectrum. To assess any time variations quantitatively,
we computed $(f_1-f_2)/\sqrt{(\sigma_1^2+\sigma_2^2)}$ for each pair
of spectra, where $f_1$ and $f_2$ are the flux densities at a given
wavelength and $\sigma_1$ and $\sigma_2$ are their uncertainties. Plots
of this quantity versus wavelength revealed that the only variable
regions in the spectra (including regions where absorption lines are
present) could be explained by variable geocoronal emission lines
or uncertainties in calibration. We also measured the average flux
densities of the three spectra in several wavelength bands which
are free from emission and absorption; these measurements are listed
in Table~\ref{fluxmeasur} and confirm that there are no variations
within the $\sim$10\% uncertainty of each measurement.  A careful look
at Table~\ref{fluxmeasur} suggests that the flux of the 2002 Mar 29
spectrum may be somewhat lower (by $\sim$~3\%) than for the two spectra
taken about a year later, especially longward of $\sim 990$~\AA.
However, this possible variation is not highly significant ($\la
1\sigma$). Since we found the spectra from the three epochs to be
consistent, we combined them into one 92.2~ks spectrum to increase
the S/N of our analysis (see Figure~\ref{fusespec}).

\begin{deluxetable}{cccc}
\tablecolumns{4}
\tablewidth{0pt}
\tablecaption{Average Flux-Density Measurements
\label{fluxmeasur}}
\tablehead{
\colhead{Wavelength range\tablenotemark{a}} &
\colhead{2002 Mar 29\tablenotemark{b}} &
\colhead{2003 Jan 18\tablenotemark{b}} &
\colhead{2003 Mar 19\tablenotemark{b}}
 }
\startdata
 926.5--927.2 & $13.39\pm1.22$ & $13.86\pm1.21$ & $13.37\pm1.23$ \\
 939.8--941.8 & $11.01\pm0.76$ & $10.95\pm0.70$ & $10.29\pm0.73$ \\
 956.9--958.0 & $13.09\pm1.09$ & $12.61\pm0.98$ & $13.18\pm1.07$ \\
 967.1--968.2 & $\phn9.28\pm1.10$ & $10.69\pm1.02$ & $\phn8.92\pm1.05$ \\
 998.2--999.2 & $13.93\pm1.04$ & $16.81\pm0.94$ & $15.24\pm1.01$ \\
1019.2--1020.7 & $15.45\pm0.72$ & $16.92\pm0.64$ & $16.56\pm0.69$ \\
1030.1--1031.6\tablenotemark{c} & $\phn7.81\pm0.62$ & $\phn8.31\pm0.56$ & $ \phn8.15\pm0.59$ \\
1042.0--1043.0\tablenotemark{d} & $21.38\pm0.90$ & $21.69\pm0.79$ & $21.33\pm0.87$ \\
1056.1--1057.6 & $12.76\pm0.76$ & $14.36\pm0.67$ & $13.16\pm0.71$ \\
1069.0--1070.0 & $13.29\pm0.97$ & $14.42\pm0.85$ & $14.50\pm0.92$ \\
1089.0--1090.0 & $12.22\pm1.65$ & $13.33\pm1.33$ & $11.29\pm1.34$ \\
1095.0--1096.5 & $\phn9.98\pm0.87$ & $11.88\pm0.72$ & $11.69\pm0.75$ \\
1114.4--1116.4 & $10.79\pm0.37$ & $11.97\pm0.33$ & $10.99\pm0.35$ \\
1127.4--1129.4 & $\phn9.59\pm0.36$ & $11.41\pm0.33$ & $10.75\pm0.35$ \\
1145.4--1147.4 & $\phn8.97\pm0.41$ & $\phn9.17\pm0.34$ & $\phn9.50\pm0.38$ \\
1181.3--1183.3 & $13.30\pm0.96$ & $14.04\pm0.80$ & $\phn\phn13.56\pm0.88$ 
\enddata
\tablenotetext{a}{In units of \AA\ in the rest frame.}
\tablenotetext{b}{In units of $10^{-15}$ \ergcmsA .}
\tablenotetext{c}{Measurement in the trough of \ion{O}{6}~$\lambda$1032.}
\tablenotetext{d}{Measurement on the red wing of \ion{O}{6}~$\lambda$1038.}
\end{deluxetable}

We also checked for flux variations within each observation by comparing
the spectra between single orbits and by combining 4--5 orbits together.
The spectra from single orbits have very poor S/N (of order 1), and no
variability is detected between them within the limiting S/N of the data.
Analyses of the spectra made by combining 4--5 orbits together show
that any FUV variability is less than $\sim 20$\% (considering several
wavelength bands as given in Table~2). 

We compared the {\it FUSE\/} observations to the long-term {\it
RXTE\/} 2--10~keV light curve of NGC\,4051 (Lamer et~al. 2003;
McHardy et~al. 2003; P. Uttley and I. M. McHardy 2003, private
communication). During the year of the {\it FUSE\/} observations,
NGC\,4051 was observed with {\it RXTE\/} every two days allowing
fairly contemporaneous comparison of the X-ray and FUV fluxes
(although, as noted in \S\,1, NGC\,4051 is highly X-ray variable on
even shorter timescales). The fluxes from the three {\it RXTE\/}
observations closest in time to the {\it FUSE\/} observations are
consistent [the 2--10~keV fluxes are measured to lie in the range
(3--3.5)$\times 10^{-11}$ \ergcms]. We also compared the {\it FUSE\/}
observations to optical observations taken on similar dates (within
about 10 days) at the Wise Observatory. The optical observations
show no variations either. This is consistent with the results of
Peterson et al. (2000) and Shemmer et~al. (2003) who find NGC\,4051
to vary by only a few percent in the optical.

\section{Spectral Analysis}

The most prominent intrinsic features seen in the {\it FUSE\/} spectrum
(Figure~\ref{fusespec}) are the \ion{O}{6} emission and absorption
lines and the \ion{H}{1} Lyman series absorption lines, which are
detected up to the Lyman edge. We also identify weak intrinsic emission
from \ion{N}{3}, \ion{C}{3}, and \ion{He}{2}. The \ion{C}{3} line shows
absorption, while no intrinsic absorption is detected in the \ion{N}{3}
and \ion{He}{2} lines. All absorption lines are blueshifted relative
to the systemic velocity, and some show multiple absorption systems.

\begin{figure}
\centerline{\includegraphics[width=8.5cm]{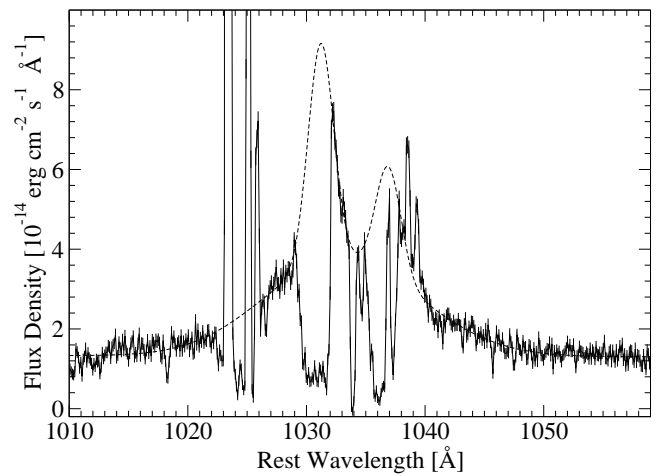}}
\caption{Fit to the \ion{O}{6} emission lines using two Gaussians for
each line (dashed curve) plotted over the {\it FUSE} spectrum (solid
curve) which is binned to $\sim 0.1$ \ \AA~pix$^{-1}$. The 1$\sigma$
uncertainties for the points are also shown. The fit parameters are
given in Table~\ref{o6fittab}.
\label{o6fit} }
\end{figure}

\begin{figure}
\centerline{\includegraphics[width=8.5cm]{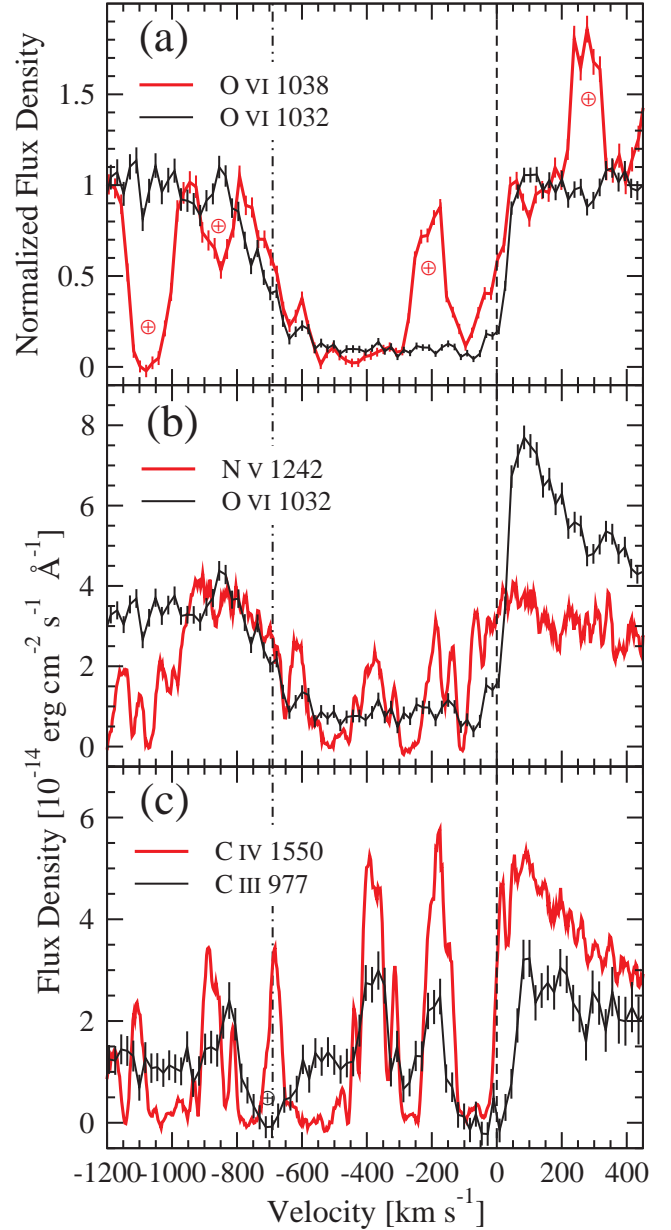}}
\vspace{-2.1cm}
\caption{
(a) Velocity spectra for the two normalized \ion{O}{6} lines.  Airglow
geocoronal emission lines and interstellar Galactic absorption lines
are marked with~$\oplus$. The dashed line marks the rest frame zero velocity
and the dotted-dashed line marks the recession velocity of NGC\,4051, i.e., absorption around this velocity is suspected to be Galactic.
(b) 
Velocity spectrum comparison between the \ion{O}{6}~$\lambda 1032$
line and the \ion{N}{5}~$\lambda 1242$ line from Collinge et
al. (2001).  The individual \ion{N}{5}~$\lambda 1242$ absorption
systems are blended into one wide trough in the \ion{O}{6}~$\lambda
1032$ line.
(c) Velocity spectrum comparison between the \ion{C}{3}~$\lambda
977$ line and the \ion{C}{4}~$\lambda 1550$ line from Collinge et
al. (2001).  The flux of the \ion{C}{4}~$\lambda 1550$ spectrum was
scaled by 0.5 for clarity of the plot. There is a good correlation
between the absorption systems seen in both ions, though the
\ion{C}{3}~$\lambda 977$ spectrum has poorer resolution. Not
all 10 absorption systems seen in \ion{C}{4}~$\lambda 1550$ can be
identified in the \ion{C}{3}~$\lambda 977$ absorption, though
the \ion{C}{3} profile is consistent with all 10 systems being present.
\vglue -0.5cm
\label{o6velspec} }
\end{figure}

To study the intrinsic \ion{O}{6} absorption, we fitted each of the two
doublet emission lines with two Gaussians, keeping the width of each
Gaussian the same for both doublet lines and keeping the flux ratio the
same as the oscillator-strength ratio (2:1). The fit results are shown
in Figure~\ref{o6fit} and listed in Table~\ref{o6fittab}. The observed
spectrum was then divided by this emission model, and the resulting
normalized spectrum was used to create the absorption velocity spectra
shown in Figure~\ref{o6velspec}a. The absorption-line profiles of
the two \ion{O}{6} lines are similar with no noticeable features
in the trough which ranges from about 50 \kms\ to $-800$ \kms. The
overall spread of velocities is in agreement with the 10 absorption
systems identified in the UV lines of \ion{C}{4} and \ion{N}{5}
(see Figure~\ref{o6velspec}b); however, no individual absorption systems
are seen in \ion{O}{6}. Since the resolution of the FUV spectrum
is high enough to resolve these systems at least in part (see
Figure~\ref{o6velspec}c), and since the troughs of the two lines have
similar depths (thus departing from the oscillator-strength ratio),
we conclude that the \ion{O}{6} absorption lines are saturated and
have broad enough damping wings to create a single trough. The
\ion{O}{6} trough is not completely black which indicates partial
covering of this absorber.\footnote{We rule out the possibility that
the trough is not black due to an instrumental scattering problem since
the Galactic absorption line at $\sim~1034$~\AA\ reaches zero flux.
We also rule out the possibility of scattering of nuclear light into
the line of sight since other absorption lines reach zero flux; see
the discussion at the end of \S\,\ref{fuvem}.}
From Figure ~\ref{o6velspec}a we estimate the absorber covering factor
to be $\sim$~0.9--0.95.

In addition to its FUV lines, the ion \ion{O}{6} is also expected
to have X-ray absorption lines at 19.135, 19.341, 21.79, and  22.01
\AA\ (E. Behar 2003, private communication; see also Pradhan 2000).
The \ion{O}{6}~$\lambda\lambda$1032, 1038 absorption systems detected
in the {\it FUSE\/} spectrum have a combined width of $\sim 700$ \kms
. The velocity resolution of the {\it Chandra} HETGS (LETGS) spectrum,
at $\sim$20 \AA , is $\sim$350 (600) \kms ; hence the \ion{O}{6}
absorption lines can be resolved by these instruments. Examination
of the HETGS and LETGS spectra shows that the data are consistent
with these lines being present, though the poor S/N prevents any
significant detection.

In Figure~\ref{o6velspec}c we compare the absorption profile of
\ion{C}{3} with that of \ion{C}{4}~$\lambda 1550$ from Collinge
et al. (2001). \ion{C}{3} shows absorption in the velocity ranges
of $-150$ to $50$, $-330$ to $-230$, and possibly around $-450$
\kms . Within the limitations of the FUV spectral resolution ($\sim
20$ \kms), the main absorption systems that were identified in the
\ion{C}{4} STIS profile are also seen in the \ion{C}{3} {\it FUSE\/}
profile.  The recession velocity of NGC\,4051 is $\sim +690$ \kms .
Any absorption at an apparent rest-frame blueshifted velocity of
$\sim -690$ \kms\ is probably Galactic or contaminated by Galactic
absorption, and no useful information about the intrinsic absorption
in NGC\,4051 can be deduced from absorption near this velocity.

\begin{deluxetable}{cccc}
\tablecolumns{4}
\tablewidth{0pt}
\tablecaption{\ion{O}{6} fit parameters
\label{o6fittab}}
\tablehead{
\colhead{Gaussian} &
\colhead{Normalization\tablenotemark{a}} &
\colhead{$\sigma$\tablenotemark{b} \ [\AA]} &
\colhead{FWHM\tablenotemark{b} \ [\kms]}
 }
\startdata
I   & 0.540  & 1.074  &      735 \\
II  & 0.179  & 5.831  &     3970 
\enddata
\tablecomments{All Gaussian centroids were set to the same velocity. In
the rest-frame spectrum of the source they correspond to 1031.22\,\AA\
and 1036.91\,\AA , i.e., the fitted \ion{O}{6} emission lines are
blueshifted by $-200$\,\kms\ relative to the optical lines.}
\tablenotetext{a}{The \ion{O}{6}~$\lambda$1032 line normalization
in units of $10^{-13}$ \ergcmsA . The \ion{O}{6}~$\lambda$1038
line normalization was set to half the \ion{O}{6}~$\lambda$1032
line normalization.}
\tablenotetext{b}{$\sigma$ and FWHM are the same for the two lines
of the doublet.}
\end{deluxetable}

\begin{figure}
\centerline{\includegraphics[width=8.5cm]{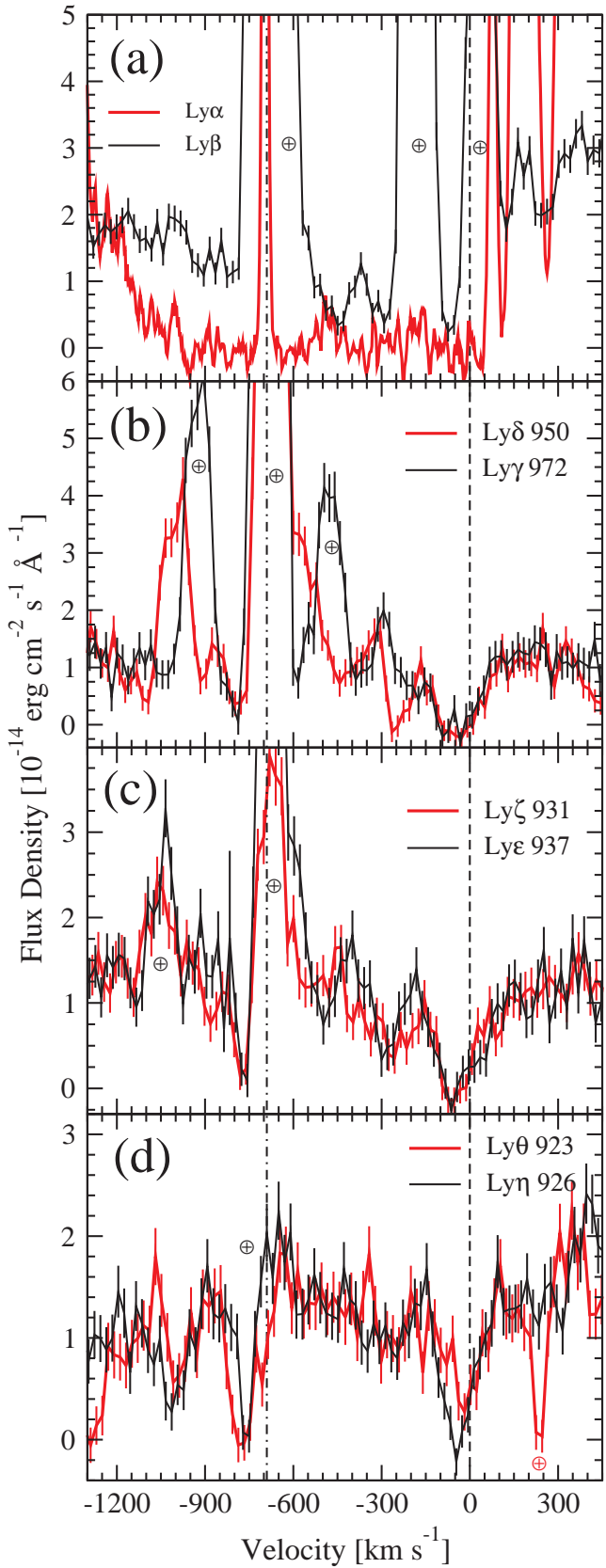}}
\caption{Velocity spectra for the first eight lines in the \ion{H}{1} Lyman
series. The vertical dashed line marks zero velocity.  Airglow
geocoronal emission lines and interstellar Galactic absorption lines
are marked with~$\oplus$. The dashed line marks the rest frame zero velocity
and the dotted-dashed line marks the recession velocity of NGC\,4051, i.e., absorption around this velocity is suspected to be Galactic.
\label{lyvelspec} }
\end{figure}

Velocity spectra for the first eight lines in the \ion{H}{1}
Lyman series are shown in Figure~\ref{lyvelspec}. The Ly$\alpha$
absorption observed with STIS (from Collinge et al. 2001) is
shown in Figure~\ref{lyvelspec}a together with Ly$\beta$ from {\it
FUSE}. Ly$\alpha$ does not show individual intrinsic systems but
rather has a broad absorption trough which reaches up to a velocity
of $\sim -1200$ \kms . All the Lyman series lines observed with {\it
FUSE} show geocoronal Lyman emission lines at $\sim -550$ to $-750$
\kms . The absorption trough of Ly$\beta$ is further contaminated
with geocoronal \ion{O}{1} lines which mask the absorption in the
range $\sim 90$ to $-270$ \kms . The Ly$\gamma$ and Ly$\delta$ lines
are shown in Figure~\ref{lyvelspec}b. Both lines are contaminated at
velocities $\la -430$ \kms .

\begin{figure*}
\centerline{\includegraphics[width=17cm]{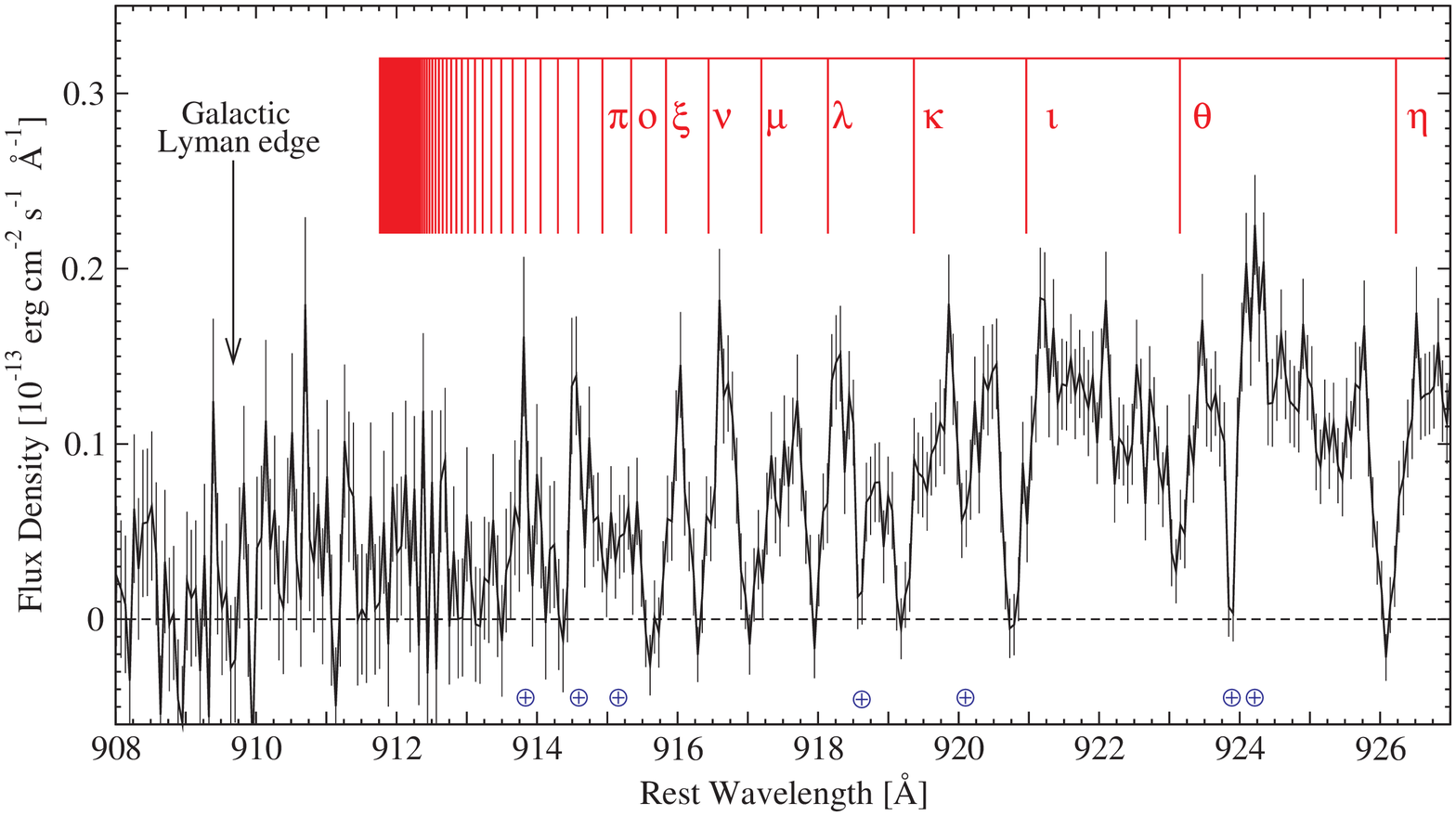}} 
\caption{The combined 92.2 ks {\it FUSE\/} spectrum of NGC\,4051
binned to $\sim 0.07$ \AA~pix$^{-1}$ showing the Lyman absorption
lines of \ion{H}{1} toward the Lyman limit. All absorption lines up
to Ly$\xi$ appear to be saturated and are blueshifted by $-50$ \kms .
The Galactic Lyman limit absorption starts at 909.7 \AA.  The dashed
horizontal line shows the zero flux level.  Airglow geocoronal
emission lines and interstellar Galactic absorption lines are marked
with~$\oplus$.
\label{lylines} }
\vglue -0.4cm
\end{figure*}

The \ion{H}{1} Lyman absorption lines are seen all the way to the Lyman
edge (see Figure~\ref{fusespec} top panel and Figure~\ref{lylines}). We
identify two main \ion{H}{1} Lyman absorption systems: one at
$-50\pm30$~\kms\ which is detected in the lines from Ly$\gamma$
to Ly$\xi$ (after which the lines are blended together and blended
with airglow emission lines; see Figure~\ref{lylines}), and one at
$-240\pm40$~\kms\ which is detected in the lines from Ly$\gamma$
to Ly$\lambda$ (after which it is weak and blended with the $-50$
\kms\ absorption lines). All lines identified in the $-50$ \kms\
system are saturated.

We used the XVOIGT software (Mar \& Bailey 1995) to estimate the
column densities of the two \ion{H}{1} Lyman absorption systems.
XVOIGT allows the interactive fitting of Voigt profiles to atomic
absorption lines in astronomical spectra. We fitted simultaneously
all identified lines in each system in order to determine the column
densities and the Doppler widths of the systems, assuming that each system
is formed by a single absorber. The large number of lines fitted
in each system allowed us to fit even the saturated lines, using the
differences in their saturated Voigt profiles. We find the $-50$ \kms\
system to have a column density of $2.2^{+1.3}_{-1.1}\times10^{17}$
\cmii\ and the $-240$ \kms\ system to have a column density of
$1.0^{+0.6}_{-0.5}\times10^{16}$ \cmii ; we note that these column
density estimates could have significant systematic errors if there is
unresolved velocity substructure in the lines. For both systems we find
a Doppler width of $b=35\pm15$ \kms . Shortward of the Lyman edge the
flux drops to about 30\% of the continuum value. This can be explained
by the total column density found above ($\sim 2.3\times10^{17}$ \cmii
) which can produce such Lyman continuum absorption. The Galactic Lyman
edge appears at 909.7~\AA\ in the NGC\,4051 rest-frame spectrum
(see Figure~\ref{lylines}), and shortward of it the spectrum is
totally absorbed.

There is probably yet another absorption system of \ion{H}{1}
at around $-400$ to $-500$ \kms .  This system is seen only
in Ly$\beta$, Ly$\delta$, Ly$\epsilon$, and possibly Ly$\zeta$
(Figure~\ref{lyvelspec}). This absorption is shallower than the other
two systems and is seen only in 3--4 lines; hence we cannot measure
its low column density as for the other two systems.

Two Ly$\alpha$ absorption systems with redshift velocities of $\sim
110$ and $\sim 260$ \kms\ were suggested by Collinge et~al. (2001) to
originate in low-ionization, high-velocity clouds within NGC\,4051
(see their Figure 9). These are also identified in Ly$\beta$
absorption and marginally identified in  Ly$\gamma$ and Ly$\delta$
(Figure~\ref{lyvelspec}a and \ref{lyvelspec}b).

In the HETGS spectrum of NGC\,4051, Collinge et~al. (2001)
found a high-velocity absorption system at $-$2340 \kms . We
also looked for this system in the {\it FUSE} spectrum. For the
\ion{O}{6}~$\lambda$1032 line this high-velocity absorption system
falls in the Ly$\beta$ absorption trough where it is blended
with geocoronal emission. For the \ion{O}{6}~$\lambda$1038
line this absorption system falls on the blue wing of the
\ion{O}{6}~$\lambda$1032 trough. Figure~\ref{o6velspec}a shows that
the blue wings of the two \ion{O}{6} troughs are similar, although
the blue wing for \ion{O}{6}~$\lambda$1032 is a bit lower than for
\ion{O}{6}~$\lambda$1038.  However, this difference is not significant
enough to claim a detection since such a small discrepancy could
be caused by calibration effects or by uncertainties in the fit to
the \ion{O}{6} emission.  We also do not detect the $-$4500 \kms\
absorption system reported by van der Meer et al. (2003) from the
{\it Chandra\/} LETGS spectrum, nor do we detect the $\sim -6500$
\kms\ or $\sim -16500$ \kms\ (depending on the interpretation of an
absorption line seen at $\sim 7.1$ keV) absorption system reported
by Pounds et al. (2003) from an {\it XMM-Newton} observation.

\section{Discussion}

\subsection{Far Ultraviolet Emission and Absorption}
\label{fuvem}

In these first FUV spectra of NGC\,4051, we detect emission from
\ion{O}{6}, \ion{N}{3}, \ion{He}{2}, and \ion{C}{3}. The \ion{O}{6}
emission lines are fitted with two Gaussians with FWHMs of $\sim 700$
and $\sim 4000$ \kms ; the overall line profiles are significantly
broader than for the optical lines.  This is in agreement with the
majority of other NLS1s which show broad wings (several thousands of
\kms ) on their high-ionization UV lines (e.g., Rodriguez-Pascual,
Mas-Hesse, \& Santos-Lleo 1997).
We also find the Gaussian centroids to be blueshifted by $-200$\,\kms\
relative to the optical lines. This is again in agreement with the
findings for other AGN (e.g., Richards et al. 2002 and references
therein).

We detect absorption on the blue wings of the \ion{O}{6} and
\ion{C}{3} emission lines, as well as absorption in the \ion{H}{1}
Lyman series. In \ion{H}{1} we identify two main absorption
systems and possibly a third at outflow velocities of $-50\pm30$,
$-240\pm40$, and $\sim -450$ \kms . These three absorption systems
are also present in \ion{C}{3} (Figure~\ref{o6velspec}c). The three
FUV systems correspond in outflow velocity to UV systems 7--8, 5,
and 2--3 which were found by Collinge et~al. (2001) in \ion{N}{5},
\ion{Si}{4}, and  \ion{C}{4}. Thus, it is plausible that each of
the three systems detected in the FUV is a blend of several narrower
systems which cannot be resolved by {\it FUSE}.\footnote{We note that
UV system 5 (and perhaps other systems) of Collinge et~al. (2001)
is saturated and broad; hence it might well be a blend of several
systems which are intrinsically blended (rather than just blended
due to instrumental resolution).}

An interesting relation is present between the strengths of the
absorption systems and their outflow velocities. The absorption system
with the largest outflow velocity ($\sim -450$ \kms ) seems to have
the weakest absorption, and it is present only in the lines up to
about Ly$\zeta$, i.e., it has a relatively low column density. The
absorption system with the smallest outflow velocity ($\sim -50$ \kms
) has the strongest absorption with relatively high column density
($2.2^{+1.3}_{-1.1}\times10^{17}$ \cmii ), and all the lines that
can be detected within the limit of the instrumental resolution
(up to Ly$\xi$) are saturated. The intermediate-velocity absorption
system (at $\sim -250$ \kms ) has an intermediate absorption depth,
detected only up to Ly$\lambda$, and it has a column density of
$1.0^{+0.6}_{-0.5}\times10^{16}$ \cmii .  This relation might be
explained if the systems with higher velocities are closer to the
central radiation source and hence are in a more intense radiation
field which yields less \ion{H}{1} in these systems.  This might
indicate that the outflow is decelerating so that its velocity
decreases as its distance from the ionizing source increases
(a decelerating outflow was also found recently in NGC\,3783;
Gabel et al. 2003).  In the lowest velocity outflow system, which
is highly saturated, we detect on the trough's wing from 100 to
$-$100 \kms\ the effect of a velocity dependent covering factor (see
Figure~\ref{o6velspec}a), i.e., the absorption profile is produced
by a different covering factor for each outflow velocity and not by
the internal Voigt profiles of the lines (e.g., Arav, Korista, \&
de Kool 2002).

We can use the multi-epoch observations to put some constraints on
the line-of-sight acceleration of the absorbers.\footnote{This might
not be the physical acceleration of the absorbing material, which
could be crossing our line of sight to the nucleus in a ``standing''
pattern.} First we use the {\it FUSE} observations which are separated
by 354.4 days in the rest frame. An upper limit on the absorption
profile difference between the two epochs is 30 \kms\ (this is the
trough location measurement uncertainty; here we used the \ion{O}{6},
\ion{C}{3}, and Ly$\gamma$ to Ly$\theta$ absorption lines). Thus, the
upper limit on the acceleration of the absorber is $9.8\times 10^{-7}$
\acc . Next we used the time between the STIS observation and the
last {\it FUSE} observation which is 1086.9 days in the rest frame.
From the observed consistency of 100 \kms\ between the two (e.g.,
Figure~\ref{o6velspec}b), we derive an upper limit on the absorber's
acceleration of $1.1\times 10^{-6}$ \acc .
The upper limit we find is at the lower range of the UV-outflow
deceleration discovered recently in NGC\,3783 of $\sim (1$--$2.5)\times
10^{-6}$ \acc\ (Gabel et al. 2003).

The \ion{O}{6} absorption we detect from NGC\,4051 is an example of
the ``smooth'' absorption morphology described in Kriss (2002). This
morphology includes objects where the \ion{O}{6} absorption is so
broad and blended  that individual \ion{O}{6} components cannot be
identified.  Another NLS1 with a ``smooth'' \ion{O}{6} absorption
morphology is Ark\,564 (Romano et~al.  2002). Other NLS1s have
``single'' morphology (like Ton\,S180 and I\,Zw\,1) or a ``blend''
morphology (like Mrk\,478; see Kriss 2002 for details). Thus, NLS1s
appear to be similar in their range of \ion{O}{6} absorption-trough
morphologies to broad-line Seyfert 1s.

The deep absorption troughs imprinted on the \ion{O}{6} and \ion{C}{3}
broad emission lines indicate these absorbers reside at a distance from
the AGN center larger than the corresponding broad line region (BLR)
size. In NGC\,4051 the BLR size has been measured, using optical Balmer
lines, to be $3\pm 1.5$ light days ($[7.8\pm3.9]\times 10^{15}$ cm;
Shemmer et~al 2003, see also Peterson et al. 2000). The BLR size for
the UV lines (higher ionization lines) is known to be smaller than
that of the optical lines (lower ionization lines) by a factor of
$\sim$ 2--3 (e.g., Peterson \& Wandel 2000).  This is attributed
to the fact that the ionizing flux density is higher closer to the
central source. Taking these considerations into account, we estimate
a lower limit for the distance of the FUV absorber from the central
continuum source to be $\sim 2\times 10^{15}$ cm.

The amount of nuclear FUV/UV light scattered around the absorbing
material and into our line of sight is practically zero since the
absorption troughs of the low-ionization ions (\ion{H}{1}, \ion{C}{3},
\ion{C}{4}, and \ion{N}{5}) reach saturation at zero flux. We find
the higher ionization ion \ion{O}{6} to have a covering factor of
$\sim$~0.9--0.95. This might indicate that the higher ionization
absorbers have a smaller covering factor than the low-ionization
absorbers.

\subsection{Ultraviolet --- X-ray Connection}

Several studies have suggested a link between the UV and X-ray
absorbers in AGN (e.g., Mathur, Elvis, \& Wilkes 1995; Crenshaw
et~al. 1999). The data presented in this paper enable us to examine
this connection in NGC\,4051 .

\begin{figure}
\centerline{\includegraphics[width=8.5cm]{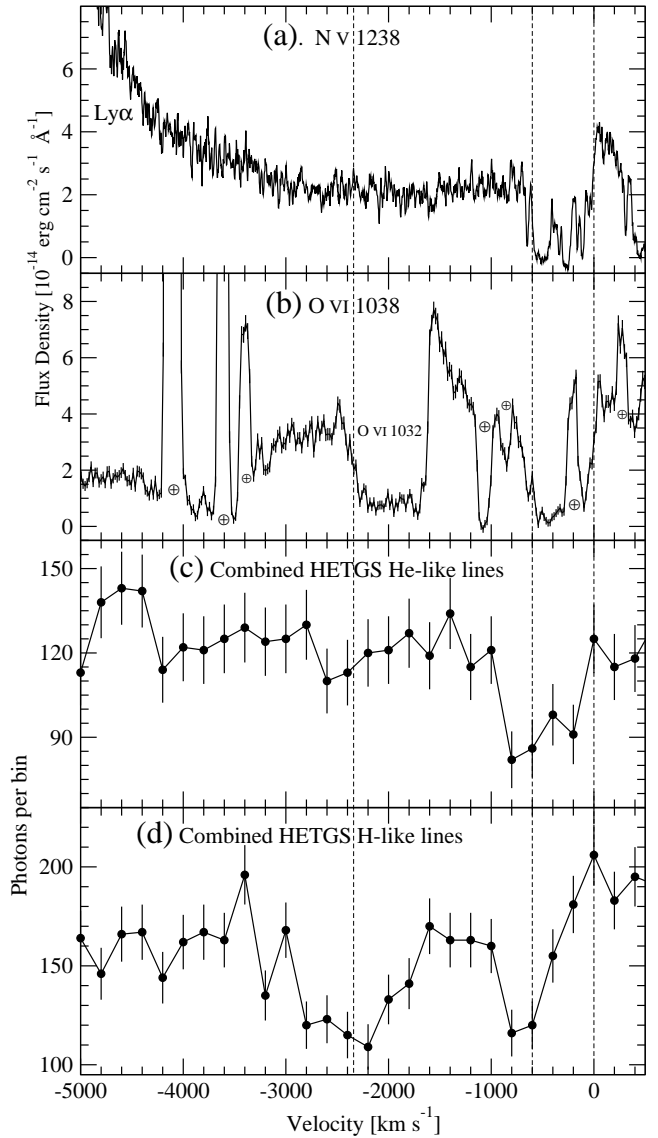}}
\caption{Line velocity profiles. 
(a) The \ion{N}{5}~$\lambda$1238 absorption from the STIS spectrum.
(b) The \ion{O}{6}~$\lambda$1038 absorption from the {\it FUSE\/} spectrum.
(c) and (d) are the combined line profiles of He-like and H-like ions,
respectively, from the {\it Chandra\/} HETGS data binned to 200 \kms\
(Figure 3 of Collinge et al. 2001).
The dashed vertical lines show 0, $-600$, and $-2340$ \kms\ velocities.
The wide absorption trough around $-2000$ \kms\ in panel (b) is
from \ion{O}{6}~$\lambda$1032.
Airglow geocoronal emission lines and interstellar Galactic 
absorption lines are marked with~$\oplus$.
\label{complines} }
\end{figure}

In Figure~\ref{complines} we compare the \ion{N}{5}~$\lambda$1238
absorption from the STIS spectrum, the \ion{O}{6}~$\lambda$1038
absorption from the {\it FUSE\/} spectrum, and combined line profiles
of He-like ions and H-like ions from the {\it Chandra\/}\,HETGS
data (Figure 3 of Collinge et~al. 2001). The resolution and S/N
of the HETGS spectrum do not allow precise measurement of the
X-ray absorption systems. The velocity shifts and FWHMs of the
FUV and UV absorption systems are consistent with the low-velocity
($-$600 \kms ) X-ray absorber which is seen in both the He-like and
H-like lines.\footnote{ We note that the profile of the low-velocity
absorber has its main trough between $\sim~ -500$ and $-900$ \kms\
in X-ray H-like lines, whereas in X-ray He-like lines it spans a wider
range from $\sim -100$ to $-900$ \kms\ (see Figures~\ref{complines}c
and \ref{complines}d). The more highly ionized absorber may be more
concentrated at higher outflow velocities, or alternatively emission
lines may be filling in the low-velocity part of the H-like absorption
trough.} Both the He-like and the H-like profiles seem to extend to
higher blueshift velocities than the FUV and UV absorbers; however,
given the HETGS resolution and the poor S/N of the X-ray spectrum,
it is not possible to determine the significance of this effect. We
do not identify in the FUV and UV absorption lines the high-velocity
absorption system (at $-$2340 \kms ) observed in the X-ray spectrum by
Collinge et~al. (2001). In particular, we do not detect this system
in \ion{O}{6}. The high-velocity absorption system in the X-ray
spectrum was detected only in the H-like ions of Si, Mg, Ne and O;
it is only marginally seen in the He-like ions of these elements
(Figure~\ref{complines}c). Thus, it is not implausible that for the
lower ionization ion \ion{O}{6} we do not detect this absorption system.
Further observations, probably with {\it Chandra\/}, are required 
to investigate the putative $-2340$~km~s$^{-1}$ absorption system further.

McHardy et al. (1995) and Nicastro et~al. (1999) modeled {\it ROSAT\/}
observations of NGC\,4051 and claimed significant ionized X-ray
absorption variability. Our observations, in contrast, fail to find any
significant FUV absorption variability (though the limited temporal
sampling might have caused a coincidence of observing on times when
conditions were similar).  This initially appears surprising if the UV
and X-ray absorbers are connected. One possible explanation is that
the low-resolution {\it ROSAT\/} data were confused by complex X-ray
spectral variability; changes in the power-law continuum and/or soft
excess could have created a false impression of ionized absorption
variability. Alternatively, this could be attributed to FUV line
saturation. Since most of the FUV absorption lines we detect appear
saturated, changes in the absorber might lead to little change in the
absorption lines. A possible exception is the \ion{C}{3} absorption
which does not seem to be heavily saturated; the lack of apparent
\ion{C}{3} absorption variability supports the idea that the absorber
does not change substantially on time scales of months.

\subsection{The Spectral Energy Distribution of NGC\,4051}

\begin{figure*}
\centerline{\includegraphics[width=17.5cm]{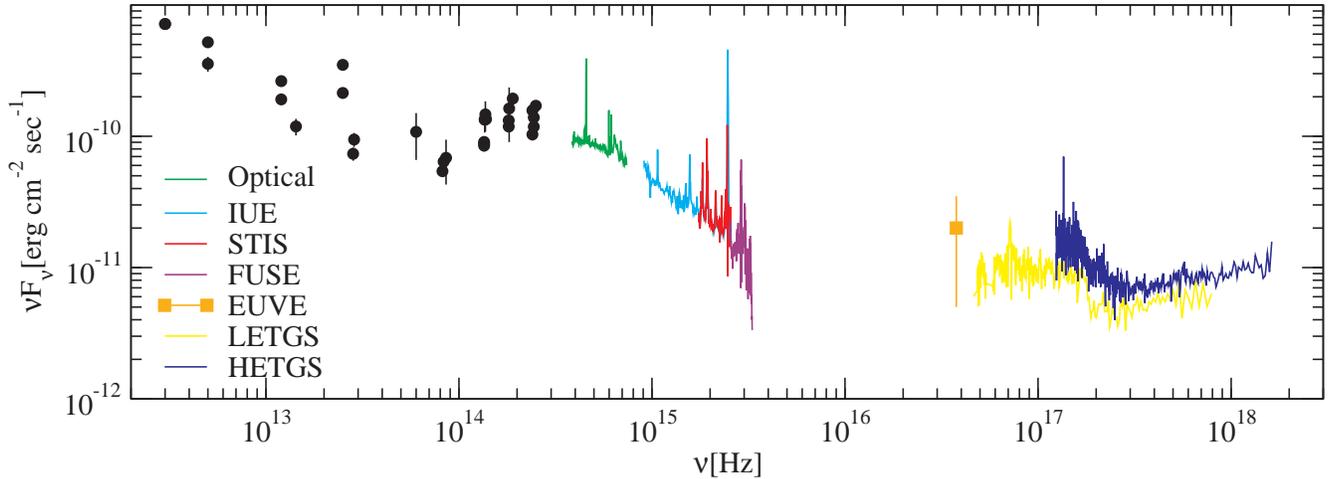}}
\caption{The SED of NGC\,4051. Optical data are the average spectrum
from Shemmer et al. (2003). {\it IUE} data are mean archive spectra
taken over $\sim 10$ yr. The {\it FUSE} data are the average spectrum
from this work. 
{\it EUVE} data are from Uttley et al. (2000), and the error bar
represents the variability range.
STIS and HETGS data are from Collinge et al. (2001);
the HETGS data represent an average of the high and low states. The
LETGS data are reported by van der Meer et al. (2003) and were taken
while NGC\,4051 was in a lower X-ray flux state.  The spectra were
heavily binned for clarity.  The infrared data were taken from the
NASA Extragalactic Database (NED) and were obtained by {\it IRAS} in
the far-IR and ground-based observatories in the near-IR. The data
were corrected for the Galactic absorption, even though it is low.
We used $E(B-V)=0.013$ mag and the extinction curve of Cardelli,
Clayton, \& Mathis (1989) for the UV to IR spectra, and $N_{\rm
H}=1.3\times 10^{20}$ \cmii\ (Elvis, Wilkes, \& Lockman 1989) with the
WABS model in XSPEC (Morrison \& McCammon 1983) for the X-ray spectra.
No intrinsic reddening correction was applied.
\label{sed} }
\vglue -0.5cm
\end{figure*}

In Figure~\ref{sed} we show a spectral energy distribution (SED) of
NGC\,4051 from the infrared to the X-ray. All data were corrected for
the Galactic absorption (see caption for details), but no intrinsic
reddening correction was applied.  In the radio region (which is
not shown in the figure) the object is radio quiet with $\nu F_\nu
\approx 10^{-16}$--$10^{-15}$ \ergcms . The data we present are not
contemporaneous, but overall most of the data represent an average over
time of the object's flux (see the figure caption for details). The
SED resembles the one presented in Komossa \& Fink (1997), although
here the SED is more detailed and has larger wavelength coverage.

The mean spectrum of NGC\,4051 from the FUV to the optical is
represented by data from four instruments and is averaged over
timescales of years. The data from all four instruments are in good
agreement (notable is the agreement of the STIS spectrum which is at a
consistent flux level with the averaged {\it IUE} spectrum observed
one-to-two decades before). The optical continuum of NGC\,4051
varies little (e.g., Shemmer et~al. 2003), the UV continuum is
also consistent with little variation, and the current work shows
that NGC\,4051 did not vary significantly in the FUV. Nevertheless,
there are notable variations in the optical line fluxes (e.g., Shemmer
et~al. 2003). Since the optical-to-FUV continuum does not vary much,
a plausible conclusion is that the line variations are driven by
the rapid and large soft X-ray and extreme UV variations and are not
driven by the near or far UV as suggested before for this object.

The observed SED of NGC\,4051 can be fitted with a power law
($f_\nu \propto \nu^{\alpha}$)  in limited spectral ranges.  We find
$\alpha$ to be $-1.46\pm 0.01$ in the optical range (5500--7500~\AA
) and $-2.03\pm 0.02$ in the UV range (1160--3000~\AA ). The
FUV cannot be fitted with a power law as it is contaminated by
emission and absorption, though Figure~\ref{sed} suggests the UV
slope continues into the FUV. In the X-ray range we find from the
HETGS spectrum in the range 2--6 keV that $\alpha = -0.84\pm0.04$.
In Table~\ref{luminosity} we give the luminosity of NGC\,4051
in several bands assuming $H_{0}=70$ km\,s$^{-1}$\,Mpc$^{-1}$,
$\Omega_{M}=0.3$, and $\Omega_{\Lambda}=0.7$.

\begin{deluxetable}{ccc}
\tablecolumns{3}
\tablewidth{0pt}
\tablecaption{Luminosity of NGC 4051 in Several Bands
\label{luminosity}}
\tablehead{
\colhead{Wavelength [\AA ]} &
\colhead{Data used} &
\colhead{Luminosity [$10^{40}$~erg\,s$^{-1}$]}  }
\startdata
4000--7500   & Optical  & 61  \\
1300--3300   & {\it IUE}      & 37  \\
1200--1700   & STIS     & 11  \\
\phn{950--1150}& {\it FUSE} & 3.8 \\
\phn{6.2--24.8} & HETGS & 16 \\
1.9--6.2 & HETGS & 13
\enddata
\tablecomments{Luminosity is computed using a cosmology of $H_0=70$
km\,s$^{-1}$\,Mpc$^{-1}$, $\Omega_{M}=0.3$, and $\Omega_{\Lambda}=0.7$,
which results in a luminosity distance of $9.8\times10^6$ pc to NGC\,4051.
\vglue -0.5cm
}
\end{deluxetable}

Overall the SED shows a gradual rise in the emitted energy from the
hard X-rays to the far-IR. Most of the observed energy from NGC\,4051
is seen in the far-IR, while the FUV emission is about an order of
magnitude weaker. The UV and FUV spectra do not show indications
for a big blue bump. Such an observed SED could be explained by dust
along the line of sight which reddens the optical-to-FUV spectrum and
re-emits energy in the far-IR. This would be similar to the observed
SED of Ark\,564 in which significant reddening was found (Crenshaw
et~al. 2002; Romano et~al. 2003). However, data from {\it EUVE\/}
and {\it ROSAT\/} strongly constrain the amount of cold gas along
the line of sight that could be associated with the putative dust
(e.g., McHardy et~al. 1995; Uttley et~al. 2000). One solution to this
difficulty could be to invoke an ionized absorber with embedded dust
(e.g., Brandt, Fabian, \& Pounds 1996; Kraemer et~al. 2000), although
even in this case there is little evidence at present for X-ray
absorption edges from the dust grains themselves (e.g., Komossa \&
Fink 1997).  Also, the relatively high flux in the far-IR could have
a significant contribution from the host galaxy (e.g., 
Ward et al. 1987; Elvis et al. 1994).

Further evidence against a dust-reddening interpretation comes from
consideration of the \ion{He}{2}~$\lambda 1640/\lambda 4686$ ratio.
We estimate the observed ratio to be a surprisingly high $19.2\pm
6.3$ from a STIS measurement of the \ion{He}{2}~$\lambda 1640$ line
[$(2.5\pm 0.3)\times 10^{-13}$ \ergcms; Collinge et~al. 2001] and
a ground-based measurement of the \ion{He}{2}~$\lambda 4686$ line
[$(1.3\pm 0.4)\times 10^{-14}$ \ergcms; Peterson et~al. 2000].
Such a value is not supportive of a reddening interpretation
which predicts the ratio to be $\la 9$ (compare with Crenshaw
et~al. 2002). If the continuum is reddened then the absorption at
1640~\AA\ is $\sim 10$ times larger than at 4686~\AA\ (by comparing
the observed local-continuum flux ratio in Figure~\ref{sed} to what
is seen in a normal AGN). Such absorption would be expected to produce
a \ion{He}{2}~$\lambda 1640/\lambda 4686$ ratio of $\sim 1$.
While our measured line ratio should be treated with caution
since variability can add a factor of $\sim 2$ to the uncertainty,
it still seems significantly higher than the expected ratio from
reddening. More accurate and simultaneous measurements of this and
other line ratios are needed to confirm this result.

An alternative to models with significant reddening is that, perhaps
due to its low black hole mass (Shemmer et al. 2003), NGC\,4051 may
have a high-temperature big blue bump that is not apparent even in
the FUV (compare our Figure~\ref{sed} with Figure~3 of Puchnarewicz
et~al. 1995). Discrimination between significant reddening and a
high-temperature big blue bump is challenging, and we leave this task
to a future study.
For example, it will be necessary to determine unambiguously if the
observed excess soft X-ray emission is due to a continuum soft X-ray
excess (e.g., Collinge et~al. 2001) or a relativistically broadened 
\ion{O}{8} recombination spectrum (Ogle et~al. 2004). 

\section{Summary}

Using three {\it FUSE\/} observations, we have presented the first
measurements of the FUV spectral and variability properties of the
intensively studied NLS1 NGC\,4051. Our main results are the following:

\begin{enumerate}
\item
NGC\,4051 shows no significant FUV variability in either its continuum 
or line properties during three epochs spanning about one year. 
\item
We detect FUV emission lines from \ion{O}{6}, \ion{N}{3}, \ion{C}{3},
and \ion{He}{2} as well as blueshifted absorption lines from
\ion{O}{6}, \ion{C}{3}, and \ion{H}{1}.
\item
The FUV absorption is generally coincident in velocity with the UV
absorption seen by STIS, although the lower FUV spectral 
resolution limits our ability to resolve all of the absorption
complexity seen in the UV. There is no evidence for new velocity 
components seen only in higher ionization FUV lines. 
\item
The FUV absorption appears coincident in velocity with only the
lowest velocity ($-600$~\kms ) X-ray absorption system known. The
several claimed X-ray absorption systems at higher velocities are not
detected in the FUV, although this does not rule out the existence
of these systems.
\item
The \ion{H}{1} Lyman series absorption lines are detected up to
the Lyman edge, and two main velocity systems are seen in these
lines. Column density estimates for these two systems indicate
that the \ion{H}{1} column density is lower for systems with larger
outflow velocities.  A possible third velocity system is also seen.
\item
Any line-of-sight acceleration of the FUV absorption is 
constrained to be $\la 1$~mm~s$^{-2}$.
\item
The deep absorption troughs imprinted on the \ion{O}{6} and \ion{C}{3}
broad emission lines indicate that the FUV absorbing material
resides at least $\sim 2\times 10^{15}$~cm from the central
continuum source. 
\item
Several low-ionization ions have troughs that reach saturation
at zero flux, significantly constraining the amount of nuclear
FUV/UV light scattered around the absorbing material. 
\end{enumerate}

\acknowledgments

We thank D. Chelouche, A. Laor, H. Netzer, O. Shemmer, and P. Uttley
for helpful discussions and sharing data. We are grateful for several
valuable suggestions by I. M. McHardy. We gratefully acknowledge the
financial support of NASA grant NAG5-13010, Israel Science Foundation
grant 545/00 (S. K.), NASA LTSA grant NAG5-13035 (W. N. B.), an
NDSEG Fellowship (M. J. C.), and the National Science Foundation
under grant AST0205990 (C. S. R.).  This research has made use of
the NASA/IPAC Extragalactic Database (NED) which is operated by the
Jet Propulsion Laboratory, California Institute of Technology, under
contract with NASA.


\begin{thebibliography}{}

\bibitem[Arav, Korista, \& de Kool(2002)]{2002ApJ...566..699A} Arav, N., 
Korista, K.~T., \& de Kool, M.\ 2002, \apj, 566, 699 

\bibitem[Brandt, Fabian, \& Pounds(1996)]{1996MNRAS.278..326B} Brandt, 
W.~N., Fabian, A.~C., \& Pounds, K.~A.\ 1996, \mnras, 278, 326 

\bibitem[Cardelli, Clayton, \& Mathis(1989)]{1989ApJ...345..245C} Cardelli, 
J.~A., Clayton, G.~C., \& Mathis, J.~S.\ 1989, \apj, 345, 245 

\bibitem[Collinge et al.(2001)]{2001ApJ...557....2C} Collinge, M.~J.~et 
al.\ 2001, \apj, 557, 2 

\bibitem[Crenshaw et al.(1999)]{1999ApJ...516..750C} Crenshaw, D.~M., 
Kraemer, S.~B., Boggess, A., Maran, S.~P., Mushotzky, R.~F., \& Wu, C.\ 
1999, \apj, 516, 750 

\bibitem[Crenshaw et al.(2002)]{2002ApJ...566..187C} Crenshaw, D.~M.~et 
al.\ 2002, \apj, 566, 187 

\bibitem[de Vaucouleurs et al.(1991)]{1991trcb.book.....D} de Vaucouleurs, 
G., de Vaucouleurs, A., Corwin, H.~G., Buta, R.~J., Paturel, G., \& Fouque, 
P.\ 1991, Third Reference Catalogue of Bright Galaxies (New York: Springer)  

\bibitem[Elvis, Wilkes, \& Lockman(1989)]{1989AJ.....97..777E} Elvis, M., 
Wilkes, B.~J., \& Lockman, F.~J.\ 1989, \aj, 97, 777 

\bibitem[Elvis et al.(1994)]{1994ApJS...95....1E} Elvis, M.~et al.\ 1994, 
\apjs, 95, 1 

\bibitem[Feldman et al.(2001)]{2001JGR...106.8119F} Feldman, P.~D., Sahnow, 
D.~J., Kruk, J.~W., Murphy, E.~M., \& Moos, H.~W.\ 2001, \jgr, 106, 8119 

\bibitem[Gabel et al.(2003)]{2003ApJ...595..120G} Gabel, J.~R.~et al.\ 
2003, \apj, 595, 120 

\bibitem[Hubble(1926)]{1926ApJ....64..321H} Hubble, E.~P.\ 1926, \apj, 64, 321 

\bibitem[Komossa \& Fink (1997)]{kf1997} Komossa, S., \& Fink H. 1997,
\aap, 322, 719

\bibitem[Kraemer, George, Turner, \& Crenshaw(2000)]{2000ApJ...535...53K} 
Kraemer, S.~B., George, I.~M., Turner, T.~J., \& Crenshaw, D.~M.\ 2000, 
\apj, 535, 53 

\bibitem[Kriss(2002)]{2002xsac.conf..109K} Kriss, G.\ 2002, in X-ray 
Spectroscopy of AGN with Chandra and XMM-Newton, eds. Boller, T., Komossa, S.,
Kahn, S., Kunieda, H., \& Gallo, L. (Garching: MPE), p. 109 

\bibitem[Lamer, McHardy, Uttley, \& Jahoda(2003)]{2003MNRAS.338..323L} 
Lamer, G., McHardy, I.~M., Uttley, P., \& Jahoda, K.\ 2003, \mnras, 338, 
323 

\bibitem[Mar \& Bailey(1995)]{1995PASA...12..239M} Mar, D.~P.~\& Bailey, 
G.\ 1995, Publications of the Astronomical Society of Australia, 12, 239 

\bibitem[Marshall, Holt, Mushotzky, \& Becker(1983)]{1983ApJ...269L..31M} 
Marshall, F.~E., Holt, S.~S., Mushotzky, R.~F., \& Becker, R.~H.\ 1983, 
\apjl, 269, L31 

\bibitem[Mathur, Elvis, \& Wilkes(1995)]{1995ApJ...452..230M} Mathur, S., 
Elvis, M., \& Wilkes, B.\ 1995, \apj, 452, 230 

\bibitem[McHardy et al.(1995)]{1995MNRAS.273..549M} McHardy, I.~M., Green, 
A.~R., Done, C., Puchnarewicz, E.~M., Mason, K.~O., Branduardi-Raymont, G., 
\& Jones, M.~H.\ 1995, \mnras, 273, 549 

\bibitem[McHardy et al. (2003)]{H003} McHardy I. M., Papadakis I. E.,
Uttley P., Page M. J., \& Mason K. O., 2003, MNRAS, in press (astro-ph/0311220)

\bibitem[Moos et al.(2000)]{2000ApJ...538L...1M} Moos, H.~W.~et al.\ 2000, 
\apjl, 538, L1 

\bibitem[Morrison \& McCammon(1983)]{1983ApJ...270..119M} Morrison, R.~\& 
McCammon, D.\ 1983, \apj, 270, 119 

\bibitem[Nicastro, Fiore, Perola, \& Elvis(1999)]{1999ApJ...512..184N} 
Nicastro, F., Fiore, F., Perola, G.~C., \& Elvis, M.\ 1999, \apj, 512, 184 

\bibitem[Ogle et al.(2004)]{2004astro.ph..1173O} Ogle, P.~M., Mason, K.~O., 
Page, M.~J., Salvi, N.~J., Cordova, F.~A., McHardy, I.~M., \& Priedhorsky, 
W.~C.\ 2004, \apj, in press (astro-ph/0401173)

\bibitem[Peterson \& Wandel(2000)]{2000ApJ...540L..13P} Peterson, B.~M.~\& 
Wandel, A.\ 2000, \apjl, 540, L13 

\bibitem[Peterson et al.(2000)]{2000ApJ...542..161P} Peterson, B.~M.~et 
al.\ 2000, \apj, 542, 161 

\bibitem[Pounds et al.(2003)]{P2003} Pounds, K. A, Reeves, J. N., King, A. R.,
\& Page, K. L. 2003, MNRAS, in press (astro-ph/0310257)

\bibitem[Puchnarewicz, Mason, Siemiginowska, \& 
Pounds(1995)]{1995MNRAS.276...20P} Puchnarewicz, E.~M., Mason, K.~O., 
Siemiginowska, A., \& Pounds, K.~A.\ 1995, \mnras, 276, 20 

\bibitem[Pradhan(2000)]{2000ApJ...545L.165P} Pradhan, A.~K.\ 2000, \apjl, 
545, L165 

\bibitem[Richards et al.(2002)]{2002AJ....124....1R} Richards, G.~T., 
Vanden Berk, D.~E., Reichard, T.~A., Hall, P.~B., Schneider, D.~P., 
SubbaRao, M., Thakar, A.~R., \& York, D.~G.\ 2002, \aj, 124, 1 

\bibitem[Rodriguez-Pascual, Mas-Hesse, \& 
Santos-Lleo(1997)]{1997A&A...327...72R} Rodriguez-Pascual, P.~M., 
Mas-Hesse, J.~M., \& Santos-Lleo, M.\ 1997, \aap, 327, 72 

\bibitem[Romano et al.(2002)]{2002ApJ...578...64R} Romano, P., Mathur, S., 
Pogge, R.~W., Peterson, B.~M., \& Kuraszkiewicz, J.\ 2002, \apj, 578, 64 

\bibitem[Romano et al.(2003)]{2003astro.ph.11206R} Romano, P.~et al.\ 2003, 
ApJ, in press (astro-ph/0311206) 

\bibitem[Sahnow et al.(2000)]{2000ApJ...538L...7S} Sahnow, D.~J.~et al.\ 
2000, \apjl, 538, L7 

\bibitem[Sembach(1999)]{1999hvc..work..243S} Sembach, K.~R.\ 1999, ASP 
Conf.~Ser.~166: Stromlo Workshop on High-Velocity Clouds, 
eds. Gibson, B.~K. \& Putman, M.~E., 243 

\bibitem[Seyfert(1943)]{1943ApJ....97...28S} Seyfert, C.~K.\ 1943, \apj, 97, 28

\bibitem[Shemmer, Uttley, Netzer, \& McHardy(2003)]{2003MNRAS.343.1341S} 
Shemmer, O., Uttley, P., Netzer, H., \& McHardy, I.~M.\ 2003, \mnras, 343, 1341

\bibitem[Uttley et al.(2000)]{2000MNRAS.312..880U} Uttley, P., McHardy, 
I.~M., Papadakis, I.~E., Cagnoni, I., \& Fruscione, A.\ 2000, \mnras, 312, 
880 

\bibitem[van der Meer, Kaastra, Steenbrugge, \& 
Komossa(2003)]{2003agnc.conf..133V} van der Meer, R.~L.~J., Kaastra, J.~S., 
Steenbrugge, K.~C., \& Komossa, S.\ 2003, ASP Conf.~Ser.~290: Active 
Galactic Nuclei: From Central Engine to Host Galaxy, 133 

\bibitem[Ward et al.(1987)]{1987ApJ...315...74W} Ward, M., Elvis, M., 
Fabbiano, G., Carleton, N.~P., Willner, S.~P., \& Lawrence, A.\ 1987, \apj, 
315, 74 

\end{thebibliography}
\end{document}